\documentclass[twocolumn]{svjour3}          
\smartqed  
\usepackage{graphicx}
\usepackage{amsmath}
\usepackage{amsfonts}
\usepackage{amssymb,latexsym}
\newcommand{\be}{\begin{equation}}
\newcommand{\ee}{\end{equation}}

\begin{document}

\title{
Cosmography and the redshift drift in Palatini $f({\cal R})$ theories
}


\author{Florencia A. Teppa Pannia \and
Santiago E. Perez Bergliaffa \and
Nivaldo Manske
}


\institute{\at
              Departamento de F\'{\i}sica Te\'orica, Instituto de F\'{\i}sica,
Universidade do Estado de Rio de Janeiro, CEP 20550-013, Rio de Janeiro, Brasil. \\
              \email{f.a.teppa.pannia@gmail.com}           
}

\date{Received: date / Accepted: date}

\maketitle

\begin{abstract}

We present an application
  to 
  cosmological models in
  $f({\cal R})$ theories within the Palatini formalism
of a method 
  that combines cosmography and the explicit form of the field equations in the calculation of the redshift drift. The method yields a sequence of constraint equations which lead to limits on the parameter space of a given $f({\cal R})$-model.  
  Two particular families of $f({\cal R})$-cosmologies capable of describing the current dynamics of the universe are explored here: (i)  power law theories of the type $f({\cal R})={\cal R}-\beta /{\cal R}^n$, and (ii) theories of the form 
  $f({\cal R})={\cal R}+\alpha \ln{{\cal R}} -\beta$. The constraints on $(n,\beta)$ and $(\alpha,\beta)$, respectively, limit the 
  values to intervals that are 
  narrower than the ones previously obtained.  As a byproduct, we show that when applied to General Relativity, the method yields values of the kinematic parameters with much smaller errors that those obtained directly from observations.

\end{abstract}

\section{Introduction}
\label{intro}

The observational evidence of the accelerated expansion of the universe can be described by assuming that gravity is governed by a theory different from General Relativity (GR) at large scales and late times. One of the most studied modifications of GR are the $f({R})$ theories of gravity, which are formulated by substituting the usual Einstein-Hilbert Lagrangian density by an arbitrary function of the Ricci curvature scalar $R$ \cite{Sotiriou2010,deFelice2010,Capozziello2011,Nojiri2010,Nojiri2017}. Cosmological models built with them seek to 
describe current astronomical data without the use of the so-called {\it dark energy}, which in the scope of the standard cosmological model amounts up to $\sim 70\%$ of the total matter content of the observable universe \cite{Planck2018}.

 Within $f({R})$-theories the dynamics of the gravitational degrees of freedom is governed by the field equations derived from the minimisation of the corresponding gravitational action. Different variational principles (usually referred as formalisms) can be considered in order to get such equations, according to the role attributed to the connection $\Gamma$
 \cite{Sotiriou2010}. Among them, the Palatini formulation is based on the assumption that the metric and the connection are independent fields. In such a case, the corresponding Riemann and Ricci tensors are constructed with a connection {\it a priori} independent of the metric \cite{Olmo2011}.  
 The Einstein-Hilbert Lagrangian is then replaced by a function $f({\cal R})$, where ${\cal R}$ is defined as ${\cal R}\equiv g^{\mu\nu} {\cal R}_{\mu\nu}(\Gamma)$, 
 and
 ${\cal R}_{\mu\nu}
$
is the Ricci tensor defined in terms of the independent connection.

The dependence of the function $f$ with the scalar curvature ${\cal R}$ introduces a set of constants,
characterising a particular family of $f({\cal R})$-cosmologies. The particular form of $f$ as well as a range of suitable values for 
the constants
must be chosen taking into account several criteria, such as 
the appropriate sequence of cosmological eras \cite{Tavakol2007}, the correct dynamics of cosmological density perturbations  \cite{Uddin2007}, as well as the correct weak-field limit at both the Newtonian and post-Newtonian levels \cite{Olmo2005b}, 
the well-posedness of the Cauchy problem \cite{Salgado2006}, and the correct fit of cosmological observables \cite{Koivisto2006b,Li2007,Liao2012}.\footnote{The  Dolgov-Kawasaki instability does not take place in 
in $f({\cal R})$ theories, see \cite{Sotiriou2007b}.}

 Among the observable quantities with the potential of discriminating between $f({\cal R})$-models and those based on GR, particular attention was recently given to the time variation of the cosmological redshift $z$ due to the variation of the expansion rate of the universe, namely the {\it redshift drift} (RD). The possibility of using this observable as a test of cosmological models was first proposed by Sandage \cite{Sandage1962}, and later developed by other authors \cite{McVittie1962,Loeb1998,Uzan2008,Quercellini2010,Amendola2013}. 
 The redshift drift (a.k.a. Sandage-Loeb effect) was considered for many years of little use in the task of distinguishing cosmological models because of the difficulties associated to its measurement. However,
 this observable may be important for Cosmology, since it allows the test of the Copernican Principle \cite{Uzan2008}, and
 has the potential to distinguish different cosmological models \cite{Quercellini2010}. 
 As first discussed in 
 \cite{TeppaPannia2013}, 
  the redshift drift can be also used 
to limit the values of the 
otherwise arbitrary constants of a given  
theory of gravity
 by resorting to its series expansion in  powers of $z$. Such expansion can be computed using two different approaches. The first one is of a cosmographical type, i.e. independent of the dynamics of the subjacent theory and only based on 
  the assumed symmetries of
 the space-time. The coefficients of the series expansion depend, in this case, on the so-called cosmographical kinematic parameters defined in terms of the time derivatives of the scale factor, the values of which follow from different observations. The second approach yields a series expansion which explicitly depends on the dynamics of the scale factor through the gravitational field equations of a given theory. The subsequent term-by-term comparison
 of the
 series expansions
leads to a  sequence of constraint equations, which explicitly relate different orders of derivatives of the scale factor (through the kinematic parameters) and the function $f({\cal R})$ and its derivatives  (and, consequently, the  constants of an specific family of $f({\cal R})$-models) evaluated at $z=0$. It is worth pointing out that such a comparison does not actually depend on the actual measurements of the redshift drift. 

The constraint equations
mentioned above
can be used in two directions: (i) to get theoretical estimations of the kinematic parameters if General Relativity is assumed, and (ii) to constrain the space-parameter of particular $f({\cal R})$-models if the kinematic parameters are derived from 
independent observational data.
 In this regard, cosmography has been widely used  to distinguish between cosmological models (see for instance \cite{Wang2009,Capozziello2008,Capozziello2011b,Aviles2012,Shafieloo2012,Capozziello2014,Capozziello2014c,Piazza2015}). In particular, the application of the comparative method coming from the redshift drift series was used in \cite{TeppaPannia2013} within the metric formalism  to constrain the space parameter of $f({R})$-models. The main goal of this work is to extend the analysis of \cite{TeppaPannia2013} to
  models built with
  $f({\cal R})$ theories in the Palatini formalism. 
We shall apply the results to the particular cases of power-law gravity ($f({\cal R})={\cal R}-\beta/ {\cal R}^n$) and logarithmic gravity ($f({\cal R})=R+\alpha \ln{({\cal R})}- \beta$). 
 
 The paper is organised as follows. General considerations about the RD and the cosmographical approach from which constraints are obtained  are presented in Section~\ref{sec:cosmography}. The application of these ideas to the standard cosmological model are given in Section~\ref{sec:kin.parameters}.
 In Section~\ref{sec:Palatinigeneral}, we present the features of $f({\cal R})$-theories within the Palatini formalism and apply the method to set limits on the parameter space of the two
 above-mentioned
  $f({\cal R})$ functions,
 and we compare our results with previously reported limits. Our final remarks are presented in Section~\ref{sec:discussion}.

\section{The redshift drift}
 \label{sec:cosmography}

The redshift of 
  a photon emitted by a source at time $t$ 
  and observed at 
    time $t_{0}$, 
  is defined as follows:
\be
\label{z1}
1+z=\frac{a(t_{0})}{a(t)}\,. 
\ee
Due to the variation of the expansion rate of the universe, the redshift of a source is indeed a function of time. Then, a second photon emitted at $t'=t+\Delta t$ will have a redshift $z(t')$. This time variation of the redshift is the so-called {\it redshift drift}, and can be expressed up to first order as \cite{Loeb1998}
\be
\label{RD1}
\frac{\Delta z}{\Delta t_{0}} = \frac{\dot a(t_{0})-\dot a(t)}{a(t)}=(1+z)H_{0}-H(z)\,,
\ee
where $\Delta t_{0}$ is the time delay between the two observed photons. Note that this observable depends
neither on specific features of the source (such as its absolute luminosity) nor on the definition of a standard ruler.

We shall present next the
two approaches to the redshift drift, each of which involve
a series expansion of this observable in terms of $z$. The first one is based on geometric and kinematic properties of the metric, namely a cosmographic treatment, whereas the other takes into account the dynamics imposed by a chosen theory of gravity through the cosmological field equations. 
 \paragraph*{A cosmographic approach.} 
The geometric and kinematic properties of the metric (\ref{dsFLRW}) are characterised by the so-called kinematic parameters, defined as the coefficients of the series expansion of the scale factor around $t_0$. In particular,
\be
H_0\equiv \left.\left(\frac{\dot{a}}{a}\right)\right|_{t_0},\ q_0\equiv - \frac{1}{H_0^2} \left.\left(\frac{\ddot{a}}{a}\right)\right|_{t_0},\ j_0\equiv  \frac{1}{H_0^3} \left.\left(\frac{\dddot{a}}{a}\right)\right|_{t_0}\,, \nonumber
   \ee
  where a dot indicates derivatives w.r.t. time,
  and 
the sub-index $0$ indicates that all quantities are evaluated at $z=0$. The aim of the cosmographic approach
  in this setting is then to compute the redshift drift in terms of these quantities.
  \footnote{The same approach has been applied to the luminosity distance in \cite{Chiba1998,Visser2005}.}

By performing a series expansion of $H(z)$ in terms of the above defined kinematic parameters, the redshift drift can be written as \cite{TeppaPannia2013}
\be
\label{RDcos}
\frac{\Delta z}{\Delta t_0}(z)= -H_0 q_0 z
+\frac{1}{2}H_0\left( q_0^2-j_0 \right) z^2
+ {\cal O}(z^3)\,.
\ee  
The coefficients of this equation, which follows from the cosmographic approach, depend purely on the properties of the FLRW metric, in the sense that no dynamical evolution for the scale factor was assumed.

 \paragraph*{A dynamical approach.}
 We shall compute the dynamical counterpart of Eq.~(\ref{RDcos}), now considering the dynamics obeyed by the scale factor $a(t)$. Let us consider again the general expression for the redshift drift given by Eq.~(\ref{RD1}). The dynamics enters through 
the Hubble parameter function, $H(z)$, the evolution of which is completely determined by the gravitational field equations once a theory of gravity is chosen.
Using the expansion of $H(z)$ in a Taylor series of the redshift, together with the chain rule ${\rm d}H/{\rm d}z = {\rm d}H/{\rm d}t \cdot {\rm d}t/{\rm d}z$,  we obtain
\begin{eqnarray}
\label{RDdyn}
&&\frac{\Delta z}{\Delta t_0}(z) = \left(H_0+\frac{\dot{H_0}}{H_0}\right) z  + \nonumber \hspace{4.cm}\\
&& \hspace{1.8cm}  + \left(\frac{\ddot{H}_0}{H_0^2}-\frac{\dot{H}_0^2}{H_0^3} 
  +\frac{\dot{H}_0}{H_0} \right) \frac{z^2}{2} +  {\cal O}(z^3)\,.
\end{eqnarray}
Note that Eq.~(\ref{RDdyn}) 
for the redshift drift 
involves the gravitational field equations through $H$ and its derivatives.

\section{Cosmography for the $\Lambda$CDM-model}
\label{sec:kin.parameters}

Cosmography is a useful standpoint to study assumptions of cosmological models which are based entirely on the Cosmological Principle \cite{Weinberg1972},
 and provides valuable model-independent information about the evolution of the scale factor and its derivatives (see also \cite{Busti2015,Dunsby2016,delaCruzDombriz2016} for an extended discussion). 
 This mathematical framework is inherently kinematic in the sense that 
 it relies only on geometrical assumptions for the metric and is independent of the dynamics obeyed by the scale factor. 
 
 We will focus in this section on the cosmographic information provided by the redshift drift in the case of the $\Lambda$CDM cosmology. The term-by-term
 comparison of both series for different powers of $z$ leads to a sequence of equations (actually, an infinite number of them) for the kinematic parameters
 in terms of the cosmological parameters. The first members of the sequence are
 \begin{eqnarray}
   \label{eq:q0GR}
   &&q_0=\frac{1}{2}\Omega_{m,0}-\Omega_{\Lambda,0}\,, \\
   \label{eq:j0GR}
   &&j_0=\frac{5}{2}\Omega^2_{m,0}-\Omega_{m,0}\Omega_{\Lambda,0}+\Omega_{\Lambda,0}^2-3\Omega_{m,0}\,.
   \end{eqnarray}
     These relations allow to estimate values for the kinematic parameters as a function of the value of $\Omega_{m,0}$. This estimation has the advantage of being independent of the restrictions associated to the convergence and the truncation of the Taylor series usually implemented in cosmography
     (see discussion in \cite{Cattoen2007,Capozziello2011b}). Taking
     for the dimensionless energy densities the values
 $\Omega_{m,0}=0.315 \pm 0.007$ and $\Omega_{\Lambda,0}=0.6847 \pm 0.0073$ \cite{Planck2018}, we obtain
     \begin{eqnarray}
     \label{q0LCDM}
       &&q_0=-0.5272 \pm 0.0081  \\ 
       &&j_0=-0.4438 \pm 0.0167\,. 
      \end{eqnarray}
The above values are in agreement with
other estimations of the kinematic parameters coming from different data sets \cite{Zhou2016,Aviles2017,Capozziello2018},
but present
significantly smaller errors.
This 
is particularly convenient 
for the case of higher-order parameters, such as $j_0$, which have large errors when estimated by other methods.

\section{Palatini $f({\cal R})$-cosmologies}
\label{sec:Palatinigeneral}
\subsection{The Palatini formalism}
\label{sec:Palatini}

Let us consider a general action 
with the form
  \begin{equation}
    \label{action}
S[g, \Gamma, \psi_m]= \int {\rm d}^4x\sqrt{-g}\left[\frac{1}{2\kappa^2}f({\cal R})+{\cal L}_{m}(g,\psi_m)\right]\,,
\end{equation}
  where $f({\cal R})$ is an arbitrary function of the curvature scalar ${\cal R}\equiv g^{\mu\nu} {\cal R}_{\mu\nu}(\Gamma)$, with ${\cal R}_{\mu\nu}(\Gamma)$ defined as ${\cal R}_{\mu\nu}(\Gamma) = -\partial_{\mu}\Gamma^{\lambda}_{\lambda\nu} + \partial_{\lambda}\Gamma^\lambda_{\mu\nu} + \Gamma^\lambda_{\mu\rho}\Gamma^{\rho}_{\nu\lambda} - \Gamma^\lambda_{\nu\rho} \Gamma^\rho_{\mu\lambda}$. 
   The matter Lagrangian density $\cal{L}_{\rm m}$ depends on the matter fields $\psi_m$, the metric $g$ and its first derivatives, but does not depend on the affine connection $\Gamma$, which appears only in the gravitational action.
  The energy-momentum tensor is conserved since the total (gravitational plus matter) action is diffeomorphism-invariant, and gravity and matter are minimally coupled by assumption \cite{Wald1984,Koivisto2006}. The Palatini $f({\cal R})$ gravity is then a metric theory (in the sense that the matter is minimally coupled to the metric and not coupled to any other fields), and hence the energy-momentum tensor $T_{\mu\nu}$ and its conservation laws will remain the ones of GR \cite{Sotiriou2010}.

The variation of the action with respect to the metric and the connection yields, respectively, \cite{Sotiriou2010} 
\be
\label{EqMetric}
f'({\cal R}) {\cal R}_{\mu\nu} - \frac{1}{2} f({\cal R}) g_{\mu\nu} = \kappa^2 T_{\mu\nu}\,, 
\ee
\begin{eqnarray}
  \label{EqGamma}
 \bar{\nabla}_ \rho\left[\sqrt{-g}\left(\delta^\rho_\lambda f'({\cal R})g^{\mu\nu}
    -\frac{1}{2}\delta^\mu_\lambda f'({\cal R}) g^{\rho\nu} \hspace{0.85cm}
    \right.\right. 
    \\ \nonumber 
   \left.\left.
    -\frac{1}{2}\delta^\nu_\lambda f'({\cal R}) g^{\mu\rho} \right)\right]= 0\,, 
\end{eqnarray}
where $f'\equiv{\rm d} f/{\rm d} {\cal R}$, and $\bar{\nabla}_\rho$ represents the derivative operator associated to the independent connection $\Gamma^\rho_{\mu\nu}$, which is assumed symmetric (torsion-less). We use units such $c=1$ and $\kappa^2=8\pi G$.

 The trace of Eq.~(\ref{EqMetric}) yields an algebraic equation for the Palatini Ricci scalar ${\cal R}$ in terms of the trace of the energy-momentum tensor $T$,
\begin{equation}
  \label{EqTraceM}
  f'({\cal R}){\cal R}-2f({\cal R})=\kappa^2 T\,.
\end{equation}
 The trace of Eq.~(\ref{EqGamma}), written as
\begin{equation}
  \label{EqTaceG}
  \bar{\nabla}_ \rho\left(\sqrt{-g}f'({\cal R})g^{\mu\nu}\right)=0\,,
  \end{equation}
is used to define the
connection of the metric $h_{\mu\nu}\equiv f'({\cal R})g_{\mu\nu}$, conformal to $g_{\mu\nu}$. Then, $\bar{\nabla}_\rho h^{\mu\nu}=0$ and the connection is actually the Christoffel symbols of the conformal metric $h^{\mu\nu}$. 
  The General Relativity case is recovered when $f'({\cal R})=
  1$ (see Eq.\ref{EqTaceG}).

Let us consider cosmological solutions by assuming the homogeneous and isotropic flat FLRW metric, described by the line element 
 \begin{equation}
   \label{dsFLRW}
{\rm d}s^2=-{\rm d}t^2+a(t)[{\rm d}r^2+r^2{\rm d}\Omega^2]\,.
 \end{equation}
The modified Friedmann equation for a dust-dominated universe then becomes \cite{Tavakol2007}
\begin{equation}
  \label{Palatini_H(z)}
  H^2=\frac{1}{6f'({\cal R})}\frac{2\kappa^2\rho_m+{\cal R}f'({\cal R})-f({\cal R})}{\left[1-\frac{3}{2}\frac{f''({\cal R})\left({\cal R}f'({\cal R})-2f({\cal R})\right)}{f'({\cal R})\left({\cal R}f''({\cal R}) -f'({\cal R})\right)}\right]^2}\,.
 \end{equation}

\subsection{Constraints on particular $f({\cal R})$-models}
\label{sec:constraints}
Two different approaches to compute a series expansion of the redshift drift were presented in the Section II, namely the cosmographic and dynamical treatments. As exemplified by the $\Lambda$CDM model, the
term-by-term comparison of both series expansion 
leads to a sequence of constraint equations to be satisfied for a given cosmological model. In the case of $f({\cal R})$-models these equations relate kinematic quantities with the parameters of the $f({\cal R})$ function. Therefore, in this section we will consider observational values for the kinematic parameters (already estimated in the literature from different data sets) to constrain the parameter space of the $f({\cal R})$ functions.

From the
comparison of the linear terms in $z$ in both expansion series (\ref{RDcos}) and (\ref{RDdyn}), 
we obtain a relation of the form
  \be
  \label{RDconstraint}
     {\cal G}(q_0,\Omega_{m,0};{\cal R}_0,f_0,f_0',f''_0, f'''_0)=0\,,
     \ee
with ${\cal G}$ a lengthy algebraic function of its arguments\footnote{Additional constraints can be obtained by taking into account higher-order terms of both series, which involve higher order derivatives of the scale factor (that is, additional kinematic parameters). See for instance \cite{PerezBergliaffa2006}.}. This expression, together with the trace Eq.~(\ref{EqTraceM}), directly implies a constraint on the space-parameter of a given $f({\cal R})$. 
Note that the absence of $H_0$ in the constraint equation \ref{RDconstraint} reduces the sources of error. 
We shall work with the values $q_0=-0.57 ^{+0.10}_{-0.08}$, estimated in \cite{Aviles2017} using the supernova type Ia JLA compilation \cite{JLA2014}, and $\Omega_{m,0}=0.315\pm 0.007$ from the last results of Planck Collaboration \cite{Planck2018}.

We present next the analysis of two $f({\cal R})$-cosmological models: (i) power-law type models, characterized by the function $f({\cal R}) = {\cal R} - \beta/{\cal R}^n$, and (ii) logarithmic type models of the form $f({\cal R})={\cal R}+\alpha \ln{{\cal R}} -\beta $. These particular choices of the function $f({\cal R})$ allow the occurrence of three cosmological phases (radiation-, matter-, and de Sitter-dominated eras) \cite{Tavakol2007}, and describe the accelerated expansion
without the introduction of non-standard sources of matter. 
Moreover, these theories satisfy the general criteria for cosmological viability discussed in Sect.~\ref{intro}.\footnote{In addition to these models, the cosmology of the so-called exponential gravity, described by the function $f({\cal R})={\cal R}-\alpha n H_0^2\left(1-{\rm e}^{-{\cal R}/(\alpha H_0^2)}\right)$, has been widely studied within the Palatini formalism \cite{Campista2011}. However we have verified that this type of models does not satisfy the necessary condition imposed by the correct sequence of cosmological eras.}

{\bf It is important to remark that both models have been shown to accommodate a phase of late-time accelerated expansion also by the integration of the corresponding equations of motion, and for a range of parameters compatible with our findings. Model (i) presents a transition from non-accelerated to accelerated expansion at around $z=1$, as shown in \cite{Cao2018,Zhai2011} by means of the evolution of the effective equation of state parameter $w_{eff}$ in terms of $z$, and also in \cite{Pires2010,Zhai2011} by studying the behaviour of $q(z)$. The same can be said about model (ii), for which the function $w_{eff}(z)$ was analyzed in \cite{Cao2018,Zhai2011}, and the deceleration factor $q(z)$ was presented in \cite{Zhai2011}.}

\subsubsection{Power law gravity: $f({\cal R}) = {\cal R}-\beta/ {\cal R}^n$}
%
Theoretical and observational studies 
determining the viability of power-law type gravity within the Palatini formalism have been presented by many authors 
 \cite{Vollick2003,Flanagan2004,Dominguez2004,Olmo2004,Capozziello2006}.
 We restrict our analysis here to the function $f({\cal R}) = {\cal R}-\beta/ {\cal R}^n$, where $n$ is a dimensionless parameter and $\beta$ is reported in units of $H_0^{2(n-1)}$. Using dynamical analysis, it was shown that this type of theories can reproduce the sequence of radiation-dominated, matter-dominated and de-Sitter eras for $ n>-1$ and $\beta > 0$ \cite{Tavakol2007}. Concerning the cosmology described by these models, a study of the evolution of the density perturbations has been presented in \cite{Uddin2007}.
 Numerical studies of the dynamic of flat models were also presented in \cite{Muller2015}.
 
 While constraints from the combination of large-scale structure, CMB data and SNIa observations could reduce the allowed space-parameter to a small region around the $\Lambda$CDM cosmology 
 ($\beta~  \approx 10^{-3}$) \cite{Koivisto2006,Li2007}, 
 other analyses combining CMB and BAO data yield 
  $(n,\beta)=(0.027^{+0.42}_{-0.23}, 4.63^{+2.73}_{-10.6})$ \cite{Tavakol2007}. These values were later improved with combinations of CMB, SNIa and BAO data, along with Hubble parameter estimations \cite{Amarzguioui2006,Carvalho2008,Santos2008}, as well as data coming from strong lensing analysis \cite{Yang2009,Liao2012}. Cosmographic constraints were obtained from the examination of the deceleration parameter \cite{Pires2010} and the so-called statefinder diagnostic \cite{Cao2018}. Recent constraints coming from cosmological standard rulers in radio quasars have been reported with considerably smaller ranges:
   $(n,\beta)=(0.052^{+0.077}_{-0.071}, 4.736^{+0.882}_{-0.681})$ \cite{Xu2017}. 

\begin{figure}
\begin{center}
  \includegraphics[angle=-90,width=0.45\textwidth]{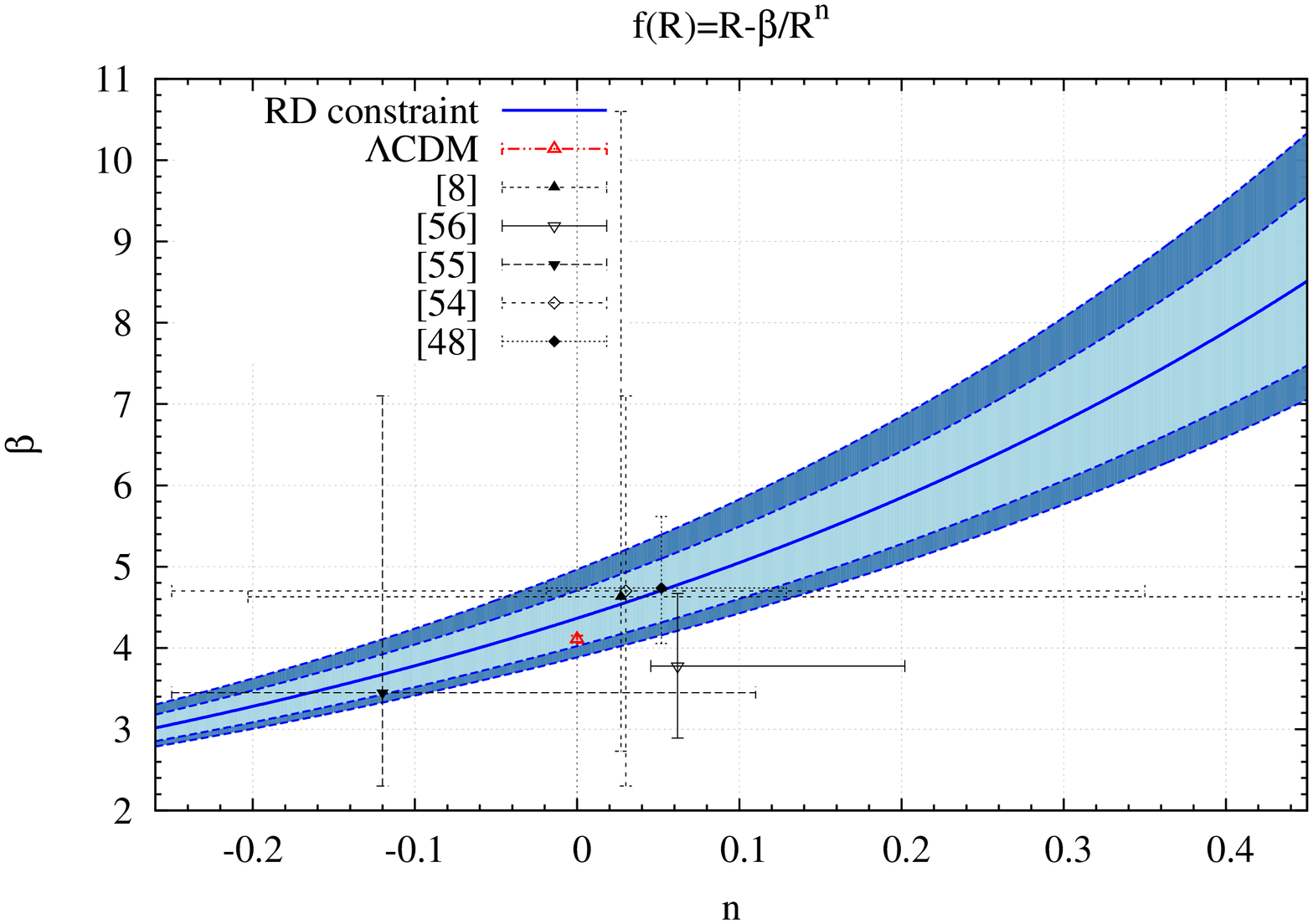}
  \vspace{0.5cm}
 \caption{Constraint on the parameter space $(n,\beta)$ for theories of the type $f({\cal R}) = {\cal R}-\beta /{\cal R}^n$ coming form the cosmographic approach to the redshift drift, together with the best-fit values previously reported and the $\Lambda$CDM case. The shadowed regions indicate the error propagation associated to the $\Omega_{m,0}$ and $q_0$ values, and a forecast corresponding to a measurement of $q_0$ with an error of $10\%$ of its value.}
 \label{fig:pow.law_n.vs.beta}
 \end{center}
\end{figure}

 In this work we present the constraint relation for the $(n,\beta)$ space, which must be satisfied by the parameters to be consistent with the cosmographic and dynamical approaches of the redshift drift. The restrictive condition coming from Eq.~(\ref{RDconstraint}) is shown in Fig.~\ref{fig:pow.law_n.vs.beta}, together with the best-fit values previously reported and
 their associated error bars.
  The shadowed regions indicate the error propagation associated to the $\Omega_{m,0}$ and $q_0$ values. Since the reported error for the $q_0$ estimation is large, we also include a forecast corresponding to a measurement of $q_0$ with an error of $10\%$ of its value. The plot shows that mildly improved estimations of the deceleration parameter would narrow even more the allowed region on the parameter space. Our constraint relation 
  is consistent with the $\Lambda$CDM model, which is recovered for $(n,\beta)=(0, 4.1082 \pm 0.0438)$ according to the latest CMB data \cite{Planck2018}, and 
  is in good agreement with other limits for this model. Furthermore, note that the 
  error associated to our constraint
  is less than that of all but one of the 
  best-fit values, and one of these is discarded by the constraint. 

 \subsubsection{Logarithmic gravity: $f({\cal R}) = {\cal R}+ \alpha \ln{{\cal R}} -\beta $}
 \label{subsec:log}

 This type of theories within the Palatini formalism was firstly studied in \cite{Meng2004b}, where it was shown that logarithmic terms in the action may
 be of use in the description of the accelerated expansion 
 and reduce to the standard Friedmann evolution for high redshifts. Furthermore, it was shown that  logarithmic gravity is one of the simplest forms capable of reproducing a suitable sequence of cosmological eras \cite{Tavakol2007}.
  Limits on the parameter space $(\alpha,\beta)$, where $\alpha$ and $\beta$ are computed in units of $H_0^2$,
have been less studied than that in the previous case. Estimations using different sets of data coming from SN, BAO and CMB observations were obtained in \cite{Tavakol2007}, with best-fit values $(\alpha,\beta)=( 0.11^{+1.75}_{-1.11},4.62^{+3.54}_{-5.58})$. 
A more recent work has reported 
 $(\alpha,\beta)=(-0.48^{+1.26}_{-2.67},3.58)$ \cite{Cao2018} (see also \cite{Zhai2011}).
  
\begin{figure}
  \includegraphics[angle=-90,width=0.45\textwidth]{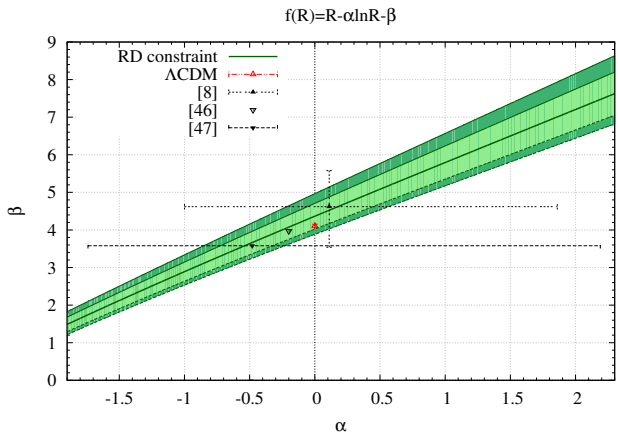} 
    \vspace{0.5cm}
  \caption{
  Constraint on the parameter space $(\alpha,\beta)$ for theories of the type $f({\cal R})={\cal R}+\alpha \ln{{\cal R}} -\beta $ coming from the cosmographic approach to the redshift drift, together with the best-fit values previously reported and the $\Lambda$CDM case. The shadowed region indicates the error propagation associated to the $\Omega_{m,0}$ and $q_0$ values, and a forecast corresponding to a measurement of $q_0$ with an error of $10\%$ of its value. }
  \label{fig:log_alpha.vs.beta}
\end{figure}

 The restriction given by Eq.~(\ref{RDconstraint}) for the case of logarithmic gravity is shown in Fig.~\ref{fig:log_alpha.vs.beta}. As in the previous analysis, the shadowed regions indicate the error propagation associated to the $\Omega_{m,0}$ and $q_0$ parameters, together with a forecast corresponding to a measurement of $q_0$ with an error of $10\%$ of its value. In this case our constraint significantly improves the limits in the parameter space and is also consistent with the $\Lambda$CDM model.

\section{Conclusion}
\label{sec:discussion}

Alternatives to General Relativity may be the key to 
describe the accelerated expansion of the universe without the use of the so-called dark energy.
In particular, cosmological models based on $f({\cal R})$-theories of gravity within the Palatini formalism have been constructed to deal with observational data sets and solar-system constraints, yielding a satisfactory description of the late time dynamics of the universe. 
 The investigation of cosmological observables capable to distinguish between 
different models can also be used as a tool to set constraints on the parameters of 
such theories. In this direction, a novel method 
for $f({\cal R})$-theories within the metric formalism was recently proposed in \cite{TeppaPannia2013}. Its essence is the comparison of two Taylor expansions of a given observable. While the first one is of the cosmographic type, namely based on the assumed symmetries of
 the space-time and independent of the dynamics obeyed by the scale factor, the second takes into account the specific dependence of $H$ with $z$, determined by the field equations of a given gravity theory. The order-by-order comparison of these two series 
 leads to a sequence of constraints relating
 the parameters of the theory. Furthermore, the method does not rely on the actual measurement of the observable, but on the condition that both series
coincide term-by-term. 
This last feature can be taken as an advantage for exploring limits imposed by yet to-be-measured quantities as is the case of the redshift drift. 
Using values for the cosmological parameters $\Omega_{m,0}$ and $\Omega_{\Lambda,0}$  obtained from observations, we have shown that a direct application of this method to GR yields theoretical estimations of the kinematic parameters 
with an error much smaller than the one that follows straight from the cosmographic approach. 

The main goal of this work was to apply the method to $f({\cal R})$-cosmological models within the Palatini formalism. Using two series expansions of the redshift drift we obtained a constraint relation, given by Eq.(\ref{RDconstraint}), on the parameter space of such models in terms of the so-called kinematic parameters. Two particular $f({\cal R})$-models were studied: (i) those based on power-law type theories, characterised by a Lagrangian function of the form $f({\cal R})={\cal R}-\beta/{\cal R}^n$, and (ii) logarithmic-gravity type models of the form $f({\cal R})={\cal R}+\alpha \ln{{\cal R}} -\beta $. 
These two particular models allow the occurrence of 
the right sequence of cosmological eras, and have been previously studied in the literature. We have used here the redshift drift expansions to impose an independent constraint relation on their parameters in terms of $q_0$ and $\Omega_{m,0}$. Our results, presented in Figs.~\ref{fig:pow.law_n.vs.beta} and \ref{fig:log_alpha.vs.beta}, are in good agreement with previously reported values, but with considerably smaller errors. These could be
reduced even more taking into account 
future improved estimations of the kinematic parameters, as shown by the forecast obtained using 
an error of ten percent of the value of $q_0$. 

To close, we would like to emphasise that the bounds obtained by the method presented here 
do not depend on the actual measurement of the redshift drift. Such bounds
can also be considered
together with those coming from theoretical considerations (such as the energy conditions \cite{PerezBergliaffa2006}), 
in order to decide whether a given $f({\cal R})$ theory furnishes an appropriate description of the current state of the universe. 

\begin{acknowledgements}
FATP acknowledges support from PNPD/CAPES and UERJ. SEPB and NM would like to acknowledge support from FAPERJ and UERJ.
\end{acknowledgements}

\bibliographystyle{spphys}       
\bibliography{bibliography}   

\begin{thebibliography}{10}
\providecommand{\url}[1]{{#1}}
\providecommand{\urlprefix}{URL }
\expandafter\ifx\csname urlstyle\endcsname\relax
  \providecommand{\doi}[1]{DOI \discretionary{}{}{}#1}\else
  \providecommand{\doi}{DOI \discretionary{}{}{}\begingroup
  \urlstyle{rm}\Url}\fi

\bibitem{Sotiriou2010}
T.P. {Sotiriou}, V.~{Faraoni}, Reviews of Modern Physics \textbf{82}, 451
  (2010)

\bibitem{deFelice2010}
A.~{de Felice}, S.~{Tsujikawa}, Living Reviews in Relativity \textbf{13}, 3
  (2010)

\bibitem{Capozziello2011}
S.~{Capozziello}, V.~{Faraoni}, \emph{{Beyond Einstein Gravity. A survey
  (...)}} (Springer Science, 2011)

\bibitem{Nojiri2010}
S.~Nojiri, S.D. Odintsov, Phys. Rept. \textbf{505}, 59 (2011)

\bibitem{Nojiri2017}
S.~Nojiri, S.D. Odintsov, V.K. Oikonomou, Phys. Rept. \textbf{692}, 1 (2017)

\bibitem{Planck2018}
N.~Aghanim, et~al.,   (2018)

\bibitem{Olmo2011}
G.J. {Olmo}, International Journal of Modern Physics D \textbf{20}, 413 (2011)

\bibitem{Tavakol2007}
S.~Fay, R.~Tavakol, S.~Tsujikawa, PRD \textbf{75}, 063509 (2007)

\bibitem{Uddin2007}
K.~{Uddin}, J.E. {Lidsey}, R.~{Tavakol}, Classical and Quantum Gravity
  \textbf{24}, 3951 (2007)

\bibitem{Olmo2005b}
G.J. {Olmo}, \prd \textbf{72}(8), 083505 (2005)

\bibitem{Salgado2006}
M.~{Salgado}, Classical and Quantum Gravity \textbf{23}, 4719 (2006)

\bibitem{Koivisto2006b}
T.~Koivisto, Phys. Rev. D \textbf{73}, 083517 (2006)

\bibitem{Li2007}
B.~{Li}, K.C. {Chan}, M.C. {Chu}, \prd \textbf{76}(2), 024002 (2007)

\bibitem{Liao2012}
K.~{Liao}, Z.H. {Zhu}, Physics Letters B \textbf{714}, 1 (2012)

\bibitem{Sotiriou2007b}
T.P. {Sotiriou}, Physics Letters B \textbf{645}, 389 (2007)

\bibitem{Sandage1962}
A.~{Sandage}, \apj \textbf{136}, 319 (1962)

\bibitem{McVittie1962}
G.C. {McVittie}, \apj \textbf{136}, 334 (1962)

\bibitem{Loeb1998}
A.~{Loeb}, \apjl \textbf{499}, L111 (1998)

\bibitem{Uzan2008}
J.P. {Uzan}, C.~{Clarkson}, G.F.R. {Ellis}, Physical Review Letters
  \textbf{100}(19), 191303 (2008)

\bibitem{Quercellini2010}
C.~Quercellini, L.~Amendola, A.~Balbi, P.~Cabella, M.~Quartin, Phys. Rept.
  \textbf{521}, 95 (2012)

\bibitem{Amendola2013}
L.~{Amendola et al.}, \jcap \textbf{12}, 042 (2013)

\bibitem{TeppaPannia2013}
F.A. {Teppa Pannia}, S.E. {Perez Bergliaffa}, \jcap \textbf{8}, 030 (2013)

\bibitem{Wang2009}
F.Y. {Wang}, Z.G. {Dai}, S.~{Qi}, \aap \textbf{507}, 53 (2009)

\bibitem{Capozziello2008}
S.~Capozziello, V.~Cardone, V.~Salzano, \prd \textbf{78}, 063504 (2008)

\bibitem{Capozziello2011b}
S.~{Capozziello}, R.~{Lazkoz}, V.~{Salzano}, \prd \textbf{84}(12), 124061
  (2011)

\bibitem{Aviles2012}
A.~{Aviles}, A.~{Bravetti}, S.~{Capozziello}, O.~{Luongo}, ArXiv e-prints
  (2012)

\bibitem{Shafieloo2012}
A.~{Shafieloo}, A.G. {Kim}, E.V. {Linder}, \prd \textbf{85}(12), 123530 (2012)

\bibitem{Capozziello2014}
S.~{Capozziello}, O.~{Luongo}, ArXiv e-prints  (2014)

\bibitem{Capozziello2014c}
S.~{Capozziello}, O.~{Farooq}, O.~{Luongo}, B.~{Ratra}, \prd \textbf{90}(4),
  044016 (2014)

\bibitem{Piazza2015}
L.~{Pizza}, \prd \textbf{91}(12), 124048 (2015)

\bibitem{Chiba1998}
T.~{Chiba}, T.~{Nakamura}, Progress of Theoretical Physics \textbf{100}, 1077
  (1998)

\bibitem{Visser2005}
M.~{Visser}, General Relativity and Gravitation \textbf{37}, 1541 (2005)

\bibitem{Weinberg1972}
S.~{Weinberg}, \emph{{Gravitation and Cosmology: Principles and Applications
  (...)}} (John Wiley \& Sons, 1972)

\bibitem{Busti2015}
V.C. {Busti}, {\'A}.~{de la Cruz-Dombriz}, P.K.S. {Dunsby},
  D.~{S{\'a}ez-G{\'o}mez}, \prd \textbf{92}(12), 123512 (2015)

\bibitem{Dunsby2016}
P.K.S. {Dunsby}, O.~{Luongo}, International Journal of Geometric Methods in
  Modern Physics \textbf{13}, 1630002-606 (2016)

\bibitem{delaCruzDombriz2016}
A.~de~la Cruz-Dombriz, PoS \textbf{DSU2015}, 007 (2016)

\bibitem{Cattoen2007}
C.~{Catto{\"e}n}, M.~{Visser}, Classical and Quantum Gravity \textbf{24}, 5985
  (2007)

\bibitem{Zhou2016}
Y.N. {Zhou}, D.Z. {Liu}, X.B. {Zou}, H.~{Wei}, The European Physical Journal C
  \textbf{76}(5), 281 (2016)

\bibitem{Aviles2017}
A.~{Aviles}, J.~{Klapp}, O.~{Luongo}, Physics of the Dark Universe \textbf{17},
  25 (2017)

\bibitem{Capozziello2018}
S.~{Capozziello}, R.~{D'Agostino}, O.~{Luongo}, MNRAS \textbf{476}(3), 3924
  (2018)

\bibitem{Wald1984}
R.M. {Wald}, \emph{{General Relativity}} (University of Chicago Press, 1984)

\bibitem{Koivisto2006}
T.~{Koivisto}, Classical and Quantum Gravity \textbf{23}, 4289 (2006)

\bibitem{PerezBergliaffa2006}
S.E. {Perez Bergliaffa}, Physics Letters B \textbf{642}, 311 (2006)

\bibitem{JLA2014}
S.~collaboration; M. Betoule~et al., \aap \textbf{568}, A22 (2014)

\bibitem{Campista2011}
M.~Campista, B.~Santos, J.~Santos, J.S. Alcaniz, Phys. Lett. \textbf{B699}, 320
  (2011)

\bibitem{Cao2018}
S.L. {Cao}, S.~{Li}, H.R. {Yu}, T.J. {Zhang}, Research in Astronomy and
  Astrophysics \textbf{18}, 026 (2018)

\bibitem{Zhai2011}
Z.X. {Zhai}, W.B. {Liu}, Research in Astronomy and Astrophysics \textbf{11},
  1257 (2011)

\bibitem{Pires2010}
N.~{Pires}, J.~{Santos}, J.S. {Alcaniz}, \prd \textbf{82}(6), 067302 (2010)

\bibitem{Vollick2003}
D.N. {Vollick}, \prd \textbf{68}(6), 063510 (2003)

\bibitem{Flanagan2004}
{\'E}.{\'E}. {Flanagan}, PRL \textbf{92}(7), 071101 (2004)

\bibitem{Dominguez2004}
A.~Dom\'{\i}nguez, D.~Barraco, Phys. Rev. D \textbf{70}, 043505 (2004)

\bibitem{Olmo2004}
G.J. {Olmo}, W.~{Komp}, ArXiv General Relativity and Quantum Cosmology e-prints
   (2004)

\bibitem{Capozziello2006}
S.~Capozziello, V.F. Cardone, M.~Francaviglia, General Relativity and
  Gravitation \textbf{38}(5), 711 (2006)

\bibitem{Muller2015}
D.~{M{\"u}ller}, V.C. {de Andrade}, C.~{Maia}, M.J. {Rebou{\c c}as}, A.F.F.
  {Teixeira}, European Physical Journal C \textbf{75}, 13 (2015)

\bibitem{Amarzguioui2006}
M.~{Amarzguioui}, {\O}.~{Elgar{\o}y}, D.F. {Mota}, T.~{Multam{\"a}ki}, \aap
  \textbf{454}, 707 (2006)

\bibitem{Carvalho2008}
F.C. {Carvalho}, E.M. {Santos}, J.S. {Alcaniz}, J.~{Santos}, \jcap \textbf{9},
  008 (2008)

\bibitem{Santos2008}
J.~{Santos}, J.S. {Alcaniz}, F.C. {Carvalho}, N.~{Pires}, Physics Letters B
  \textbf{669}, 14 (2008)

\bibitem{Yang2009}
X.J. {Yang}, D.M. {Chen}, \mnras \textbf{394}, 1449 (2009)

\bibitem{Xu2017}
T.~{Xu}, S.~{Cao}, J.~{Qi}, M.~{Biesiada}, X.~{Zheng}, Z.H. {Zhu}, ArXiv
  e-prints  (2017)

\bibitem{Meng2004b}
X.H. Meng, P.~Wang, Physics Letters B \textbf{584}(1), 1  (2004)

\end{thebibliography}

\end{document}